\documentstyle[11pt]{article}
\begin{document}

\title{\textbf{The effect of twisted magnetic field on the resonant absorption of MHD waves in coronal
loops}}
\author{K. Karami$^{1}$\thanks{E-mail: KKarami@uok.ac.ir} ,
K. Bahari${^2}$\\$^{1}$\small{Department of Physics, University of
Kurdistan, Pasdaran Street, Sanandaj, Iran}\\$^{2}$\small{Institute
for Advanced Studies in Basic Sciences (IASBS), Gava Zang, Zanjan,
Iran} }

\maketitle

\begin{abstract}
The standing quasi modes in a cylindrical incompressible flux tube
with magnetic twist that undergoes a radial density structuring is
considered in ideal magnetohydrodynamics (MHD). The radial
structuring is assumed to be a linearly varying density profile.
Using the relevant connection formulae, the dispersion relation for
the MHD waves is derived and solved numerically to obtain both the
frequencies and damping rates of the fundamental and first-overtone
modes of both the kink ($m=1$) and fluting ($m=2,3$) waves. It was
found that a magnetic twist will increase the frequencies, damping
rates and the ratio of the oscillation frequency to the damping rate
of these modes. The period ratio $P_1/P_2$ of the fundamental and
its first-overtone surface waves for kink ($m=1$) and fluting
($m=2,3$) modes is lower than 2 (the value for an untwisted loop) in
the presence of twisted magnetic field. For the kink modes,
particularly, the magnetic twists $B_{\phi}/B_z=$0.0065 and 0.0255
can achieve deviations from 2 of the same order of magnitude as in
the observations. Furthermore, for the fundamental kink body waves,
the frequency bandwidth increases with increasing the magnetic
twist.
\end{abstract}
\clearpage
\section{Introduction}
\label{intro} Transverse oscillations of coronal loops were first
identified by Aschwanden et al. (1999) and Nakariakov et al. (1999)
using the observations of Transition Region and Coronal Explorer
(TRACE). Nakariakov et al. (1999) reported the detection of spatial
oscillations in five coronal loops with periods ranging from 258 to
320 s. The decay time was $14.5\pm2.7$ min for an oscillation of
$3.9\pm0.13$ mHz. Also Wang and Solanki (2004) described a loop
oscillation observed on 17 April 2002 by TRACE in 195 $\rm\AA$. They
interpreted the observed loop motion as a vertical oscillation, with
a period of 3.9 min and a decay time of 11.9 min. Nakariakov et al.
(1999) obtained that the amplitude of the oscillations decreases by
more than 50 percent in several oscillation periods. To estimate the
energy flux of the EUV kink oscillation observed by Nakariakov et
al. (1999), we use the maximum kinetic energy flux of an oscillating
loop given by $\frac{1}{2}(\pi R^2L)\rho v_{\rm max}^2/(2\pi R
L)\tau_{\rm D}=\frac{1}{4}R\rho v_{\rm max}^2/\tau_{\rm D}$. Where
$\rho$, $v_{\rm max}$, $R$, $L$ and $\tau_{\rm D}$ are the mass
density, the peak loop velocity, the loop radius, the loop length,
and the damping time of oscillation, respectively. Taking $R=2\times
10^3$ km, $\rho=2\times 10^{-14}$ gr cm$^{-3}$ for a typical coronal
loop in active region and using $v_{\rm max}=47$ km s$^{-1}$ and
$\tau_{\rm D}=14.5$ min given by Nakariakov et al. (1999), we obtain
the energy flux as $2\times 10^4$ erg cm$^{-2}$ s$^{-1}$ which is
three orders of magnitude smaller than the radiated energy flux
$10^7$ erg cm$^{-2}$ s$^{-1}$ in active regions (see Klimchuk 2006).
Therefore, the energy flux in the EUV kink oscillations is unlikely
to be sufficient to cover heating of coronal loops.

Since the discovery of the coronal green line during the 1869
eclipse which identified as Fe XIV spectral line by Grotrian
(1939), different theories of coronal heating have been put
forward and debated (for reviews see, e.g., Hollweg 1991: Narain
and Ulmschneider 1996; Walsh and Ireland 2003; Erd\'{e}lyi 2004,
2005; Klimchuk 2006; Erd\'{e}lyi and Ballai 2007; Taroyan and
Erd\'{e}lyi 2009). Ionson (1978) was first to suggest that the
resonant absorption of MHD waves in coronal plasmas could be a
primary mechanism in coronal heating. Since then, much analytical
and numerical work has been done on the subject. Rae and Roberts
(1982) investigated both eikonal and differential equation
approaches for the propagation of MHD waves in inhomogeneous
plasmas. Hollweg (1987a,b) considered a dissipative layer
 of planar geometry to study the resonant absorption of coronal
 loops. Davila (1987) derived the resonant heating rate in the low-beta compressible resistive
 MHD approximation with the shear viscosity and found that the heating does not depend explicitly
 on the dissipation coefficients.
 Poedts, Goossens, and Kerner (1989, 1990) developed a finite element code to elaborate
 on the resonant absorption of Alfv\'{e}n
waves in circular cylinders.

Sakurai, Goossens, and Hollweg (1991a,b), Goossens, Hollweg, and
Sakurai (1992), Steinolfson and Davila (1993) did much work on
resonant absorption. Ofman, Davila, and Steinolfson (1994), and
Erd\'{e}lyi and Goossens (1994) included viscous and resistive
dissipations in their analysis and concluded that the heating rate
due to shear viscosity is comparable in magnitude to the resistive
resonant heating. Also, they concluded that the heating caused by
compressive viscosity is negligible. Goossens, Ruderman, and Hollweg
(1995) extended the analysis of Sakurai et al. (1991b) on resonant
Alfv\'{e}n waves in nonuniform magnetic flux tubes for a static MHD
equilibrium. They showed that the conservation law found in ideal
MHD is valid in dissipative MHD. They derived analytical solutions
to the dissipative MHD equations for the Lagrangian displacement and
for the total pressure perturbation. Erd\'{e}lyi and Goossens (1995)
studied the heating of solar coronal loops by resonant absorption of
Alfv\'{e}n waves in visco-resistive MHD. They pointed out that under
solar coronal conditions the two dissipative mechanisms are both
operational and that electrical resistivity appears to be slightly
more important than viscosity. Erd\'{e}lyi and Goossens (1996)
studied the effect of an equilibrium flow on resonant absorption of
linear MHD waves in compressible viscous cylindrical magnetic flux
tubes. They showed that the presence of the equilibrium flow is very
determinant for resonant absorption and significantly affect the
resonance absorption rate suffered by the incoming driving waves.
Goossens, Andries, and Aschwanden (2002) used the TRACE data of
Ofman and Aschwanden (2002) to infer the width of the inhomogeneous
layer for 11 coronal loops. Ruderman and Roberts (2002) solved the
initial value problem for a zero-beta loop driven by a kink
perturbation. They pointed out that the observed damping rate of the
coronal oscillations is due to the resonant damping of the
quasi-mode kink oscillations. They estimated the inhomogeneity
length scale (thickness of the boundary layer) of the loop with the
data of Nakariakov et al. (1999). Note that Nakariakov et al. (1999)
showed that the damping rate of the global kink oscillation observed
by the TRACE can be justified by the energy absorption due to
anomalous viscosity. Whereas Ruderman and Roberts (2002) cleared
that for large values of the viscous Reynolds number, the damping
rate is independent of the viscous Reynolds number and proportional
to the ratio of the thickness of the inhomogeneous layer and the
radius of the tube. Van Doorsselaere et al. (2004a) used the LEDA
code (Large-scale Eigenvalue solver for the Dissipative Alfv\'{e}n
spectrum) to study the resistive absorption of the kink modes of
cylindrical models. They concluded that, when the width of the
nonuniform layer was increased, their numerical results differed by
as much as 25 per cent from those obtained with the analytical
approximation.

Safari et al. (2006) studied the resonant absorption of MHD waves in
magnetized flux tubes with a radial density inhomogeneity. Using the
approximation that resistive and viscous processes are operative in
thin layers surrounding the singularities of the MHD equations, they
concluded that as the longitudinal wave number increases, the
maximum amplitude of the body eigenmodes shifts away from the
resonant layer and causes a decrease in damping rates.

An additional features of the flux tube is that of twist. There is
observational evidence for the existence of twist in the solar
atmosphere. The TRACE 171 $\rm\AA$ observations of a postflare loop
system on 1999 confirmed the existence of the twisted loops in the
corona (see Aschwanden 2005). Rotational movement along a loop
observed by Chae et al. (2000) indicates the existence of twist.
Aschwanden (2005) discusses observed noncoplanarity of loops which
implies kinked field lines which, in turn, are also intimately
related to twist.  Besides, it is unfeasible that every flux tube
within the solar atmosphere is entirely twist free in spite of the
random continuous motions observed at the footpoints (see
Erd\'{e}lyi and Carter 2006). Following Aschwanden (2005), the
magnetic twist parameter $B_{\phi}/B_z$ can be estimated as
$B_{\phi}/B_z=\tan\theta$. Here $B_{\phi}$ and $B_{z}$ are the
azimuthal and axial components of the magnetic field, respectively.
Also $\theta$ is the shear angle between the untwisted and the
twisted field line. The geometric shear angle $\theta$, can
observationally be measured in twisted coronal loops.

Bennett, Roberts, and Narain (1999) examined the influence of
magnetic twist on the modes of
 oscillations of a magnetic flux tube. They found that twist
 introduces an infinite band of body modes. Erd\'{e}lyi and Fedun (2006) studied the
wave propagation in a twisted cylindrical magnetic flux tube
embedded in an incompressible but also magnetically twisted
plasma. They found that increasing the external magnetic twist
from 0 to 0.3 caused an increase in the normalized periods of
sausage MHD waves approximately by 1$-$2$\%$. Erd\'{e}lyi and
Fedun (2007) extended the work of Erd\'{e}lyi and Fedun (2006) to
the case of compressible plasma. They found that increasing the
internal twist may result in about up to $3-5\%$ changes in
periods for sausage modes.

Erd\'{e}lyi and Carter (2006) studied the propagation of MHD waves
in a fully magnetically twisted
 flux tube consisting of a core, annulus and external region.
 They investigated their analysis by considering magnetic twist
 just in the annulus, the internal and external regions having
 straight magnetic field. Two modes of oscillations occur in
 this configuration; surface and hybrid modes. They found that
 when the magnetic twist is increased the hybrid modes cover a wide
 range of phase speeds, centered around the annulus, longitudinal
 Alfv\'{e}n speed for the sausage modes.

Carter and Erd\'{e}lyi (2007) investigated the oscillations of a
magnetic flux tube
 configuration consisting of a core, annulus and external region
 each with straight distinct magnetic field in an incompressible medium. They found that there
 are two surface modes arising for both the sausage and kink modes
 for the annulus-core model where the monolithic tube has solely one surface mode for the incompressible
 case. Also they showed that the existence and width of an annulus
 layer has an effect on the phase speeds and periods.

Carter and Erd\'{e}lyi (2008) used the model introduced by
Erd\'{e}lyi and
 Carter (2006) to include the kink modes. They found for the set of
 kink body modes, the twist increases the phase speeds of the modes. Also they showed that there are two surface modes
 for the twisted shell configuration, one due to each surface, where
 one mode is trapped by the inner tube, the other by the annulus
 itself.

Mikhalyaev and Solov'ev (2005) investigated the MHD waves
 in a double magnetic flux tube embedded in a uniform external
 magnetic field. The tube consists of a dense hot cylindrical
 core surrounded by a co-axial shell. They found two slow and two
 fast magnetosonic modes can exist in the thin double tube.

Verwichte et al. (2004), using the observations of TRACE, detected
multimode oscillations for the first time. They found that two loops
are oscillating in both the fundamental and the first-overtone
standing kink modes. According to the theory of MHD waves, for
uniform loops the ratio of the period of the fundamental to the
period of the first overtone is exactly 2, but the ratios found by
Verwichte et al. (2004) are 1.81$\pm$0.25 and 1.64$\pm$0.23.
However, these values were corrected with the improvement of the
observational error bars to 1.82$\pm$0.08 and 1.58$\pm$0.06,
respectively, by Van Doorsselaere, Nakariakov, and Verwichte (2007).
Also Verth, Erd\'{e}lyi, and Jess (2008) added some further
corrections by considering the very important effects of loop
expansion and estimated a period ratio of 1.54. All these values
clearly differ from 2. This may be caused by different factors such
as the effects of curvature (see e.g. Van Doorsselaere et al.
2004b), leakage (see De Pontieu, Martens, and Hudson 2001), density
stratification in the loops (see e.g. Andries et al. 2005;
Erd\'{e}lyi and Verth 2007; Karami and Asvar 2007; Safari, Nasiri,
and Sobouti 2007; Karami, Nasiri, and Amiri 2009), magnetic field
expansion (see Verth and Erd\'{e}lyi 2008; Ruderman, Verth, and
Erd\'{e}lyi 2008; Verth et al. 2008) and magnetic twist (see e.g.
Erd\'{e}lyi and Fedun 2006, 2007; Erd\'{e}lyi and Carter 2006;
Karami and Barin 2009).

Karami and Barin (2009) studied both the oscillations and damping of
standing MHD surface and hybrid waves in coronal loops in presence
of twisted magnetic field. They considered a straight cylindrical
incompressible flux tube with magnetic twist just in the annulus and
straight magnetic field in the internal and external regions. They
showed that both the frequencies and damping rates of both the kink
and fluting modes increase when the twist parameter increases. They
obtained that the period ratio $P_1/P_2$ of the fundamental and
first-overtone for both the kink and fluting surface modes are lower
than 2 (for untwisted loop) in presence of the twisted magnetic
field. Ruderman (2007) studied nonaxisymmetric oscillations of thin
twisted magnetic tubes in a zero-beta plasma by taking into account
the longitudinal density stratification. Using the asymptotic
analysis, he showed that the eigenmodes and eigenfrequencies of the
kink and fluting oscillations are described by a classical
Sturm-Liouville problem for a second-order ordinary differential
equation. He also concluded that the results concerning
nonaxisymmetric waves in twisted magnetic tubes obtained by Bennett
et al. (1999) for incompressible plasmas can be applied to global
nonaxisymmetric waves, i.e. kink and fluting modes, in coronal loops
despite that the coronal plasma is a low-beta plasma. This
conclusion was also in a good agreement with Erd\'{e}lyi and Fedun
(2007).

In the present work, our aim is to investigate the effect of twisted
magnetic field on the resonant absorption of standing MHD waves in
the coronal loops to justify the rapid damping of oscillations and
deviation of the period ratio $P_1/P_2$ from 2
 observed by TRACE. This paper is organized
as follows. In Section 2 we derive the equations of motion,
introduce the relevant connection formulae and obtain the
dispersion relation. In Section 3 we give numerical results.
Section 4 is devoted to conclusions.

\section{Equations of Motion}
The linearized MHD equations for an incompressible plasma are
\begin{eqnarray}
\frac{\partial\delta{\bf v}}{\partial t}=-\frac{\nabla\delta
p}{\rho}+\frac{1}{4\pi\rho}\{(\nabla\times\delta{\bf B})\times{\bf
B} +(\nabla\times{\bf B})\times\delta{\bf B}\}
+\frac{\nu}{\rho}\nabla^2\delta{\bf v},\label{mhd1}
\end{eqnarray}
\begin{eqnarray}
\frac{\partial\delta{\bf B}}{\partial t}=\nabla\times(\delta{\bf
v}\times{\bf B})+ \frac{c^2}{4\pi\sigma}\nabla^2\delta{\bf
B},\label{mhd2}
\end{eqnarray}
\begin{eqnarray}
\nabla\cdot\delta{\bf v}=0,\label{mhd3}
\end{eqnarray}
where $\delta\bf{v}$, $\delta\bf{B}$ and $\delta p$ are the Eulerian
perturbations in the velocity, magnetic field and thermal pressure,
respectively; $\rho$, $\sigma$, $\nu$ and $c$ are the mass density,
the electrical conductivity, the viscosity and the speed of light,
respectively. In the momentum Eq. (\ref{mhd1}) like Ruderman and
Roberts (2002) we write the viscous force in a simplified form
$\nu\nabla^2\delta{\bf v}$ instead of the classical Braginskii's
expression for the viscosity tensor in a magnetized plasma
(Braginskii 1965). Because according to Ruderman and Roberts (2002),
under typical coronal conditions, the coefficient of the shear
viscosity is at least 10 orders of magnitude smaller than that of
the compressional viscosity. However, in the problem of oscillations
of coronal loops, dissipation is only important in an Alfv\'{e}nic
dissipative layer embracing an ideal resonant magnetic surface.
Numerical studies by Ofman et al. (1994), and Erd\'{e}lyi and
Goossens (1994, 1995) have shown that in Alfv\'{e}nic dissipative
layers only the shear viscosity is significant, all other terms in
Braginskii's tensorial expression being neglected. Equation
(\ref{mhd3}) satisfies the incompressibility condition. The
assumption of incompressibility reduces the direct full
applicability to coronal loops except for kink modes observed by the
TRACE as they are highly incompressible perturbations (see Carter
and Erd\'{e}lyi 2007). Recently Goossens et al. (2009) showed that
in the thin tube approximation neglecting contributions proportional
to $(k_zR)^2$ then the frequency of the kink wave and its damping
due to resonant absorption are the same in the three cases including
a compressible pressureless plasma, an incompressible plasma and a
compressible plasma which allows for MHD radiation. This is expected
as kink modes are linearly incompressible. Note that here the energy
equation is absent. Because for incompressible plasmas when sound
speed (or adiabatic gas index) goes to infinity then the energy
equation is decoupled from the system of linearized MHD equations.

\begin{figure}
\includegraphics{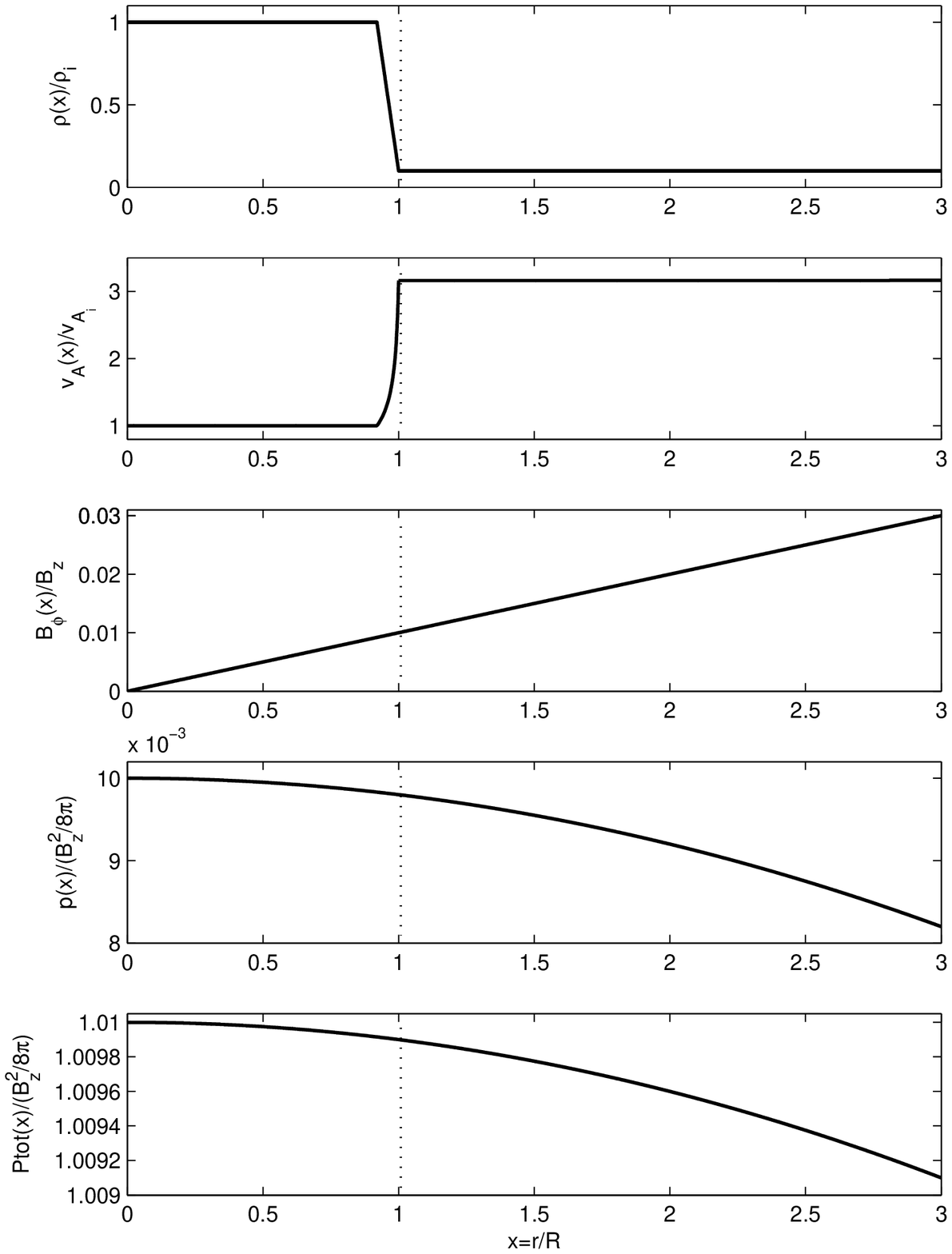}
      \vspace{13.5cm}
\caption[] {Dimensionless equilibrium quantities including density
$\rho$, Alfv\'{e}n velocity $v_{\rm A}$, azimuthal component of the
uniformly twisted magnetic field $B_{\phi}$, thermal pressure $p$,
and total pressure $p_{\rm tot}=p+B^2/(8\pi)$ against fractional
radius $x = r/R$. Auxiliary parameters are $\rho_{\rm e}/\rho_{\rm
i}=0.1$, $\frac{AR}{B_z}=0.01$,
$\beta=\frac{p_0}{B_z^2/(8\pi)}=0.01$. The parameter $\beta$ has
been taken from Smith, Tsiklauri, and Ruderman (2007).}
         \label{rho_vA_Bphi_p_Ptot}
   \end{figure}

The simplifying assumptions are as follows.
\begin{itemize}
\item The background magnetic field is assumed to be \[
\textbf{B} = \left( {0,Ar,B_z }\right),
\]
where $A$, $B_z$ are constant and the magnetic field is uniformly
twisted inside and outside the tube (see Fig.
\ref{rho_vA_Bphi_p_Ptot}). This is unphysical as
$r\rightarrow\infty$, and caution has to be exercised. Following
Ruderman (2007), to satisfy the Shafranov-Kruskal stability
criterion we assume that the azimuthal component of magnetic field
is smaller than the axial component. This assumption is also
compatible with observed weakly twisted coronal loops (see
Erd\'{e}lyi and Fedun 2006). The stability of a twisted magnetic
flux tube has been studied in some details by Bennett et al. (1999),
Ruderman (2007), and Carter and Erd\'{e}lyi (2008). Note that the
choice of $B_{\phi}=Ar$ implies that there is a constant
longitudinal current density along the flux tube.
\item From Safari et al. (2006) and Karami et al. (2009), the density profile is assumed to be
$$\rho(r)=\left\{\begin{array}{ccc}
    \rho_{{\rm i}},&(r<R_1),&\\
    \Big[\frac{\rho_{\rm i}-\rho_{\rm e}}{R-R_1}\Big](R-r)+\rho_{\rm e},&(R_1<r<R),&\\
      \rho_{{\rm e}},&(r>R),&\\
      \end{array}\right.$$
where $R$ denotes the radius of the tube and $R_1<R$ is the radius
at which the resonant absorption occurs. Also $\rho_{{\rm i}}$ and
$\rho_{{\rm e}}$ are the interior and exterior constant densities of
the tube, respectively (see Fig. \ref{rho_vA_Bphi_p_Ptot}). Note
that for resonant absorption a number of density profiles have been
considered. For instance, Ofman et al. (1994, 1995) have taken
$\rho(r)=\rho_r+(1-\rho_r)e^{-r^4}$ with $\rho_r=\rho_{\rm
e}/\rho_{\rm i}$. Here since we use the thin boundary approximation,
we assume that the inhomogeneous layer and the resonance layer
coincide (see Goossens et al. 2009). Therefore any well behavior
function of the density like that selected by Ofman et al. (1994,
1995) in the resonance layer, i.e. the region $R_1<r<R$ where the
singularity occurs, can be approximated with a linearly varying
profile.

\item The equilibrium condition, i.e.
      radial force balance equation
      $\frac{{\rm
d}}{{\rm
d}r}{\Big[}p(r)+\frac{B^2(r)}{8\pi}{\Big]}=-\frac{B^2_{\phi}(r)}{4\pi
      r}$,
      gives the equilibrium plasma pressure $p(r)$ as
\[
p(r) = p_0  - \frac{{A^2 r^2 }}{{4\pi }},
\]
where $p_0$ is the plasma pressure at the center of the tube. Since
we have no discontinuity in the background magnetic field, then the
thermal pressure remains continuous across $r=R$
 (see Fig. \ref{rho_vA_Bphi_p_Ptot}).
\item Tube geometry is a circular with cylindrical
coordinates, ($r,\phi,z$).
     \item There is no initial steady flow over the tube.
\item Viscous and resistive coefficients, $\nu$ and
$\sigma$ respectively, are constant.
\item $t$-, $\phi$- and $z$-  dependence for any of the components
$\delta{\bf{v}}$ and $\delta{\bf{B}}$ is $\exp{\{{\rm i}(m\phi+k_z
z-\omega t)\}}$. Here $k_{z}=l\pi/L$, $L$ is the length of the tube,
and $l=(1,2,\cdot\cdot\cdot)$, $m=(0,1,2,\cdot\cdot\cdot)$ are the
longitudinal and azimuthal mode numbers, respectively.
   \end{itemize}

In the remainder of this section the following steps are taken: a)
in $r < R_1$ and $r > R$, dissipative terms are neglected. Solutions
of Eqs. (1) and (2) are obtained as in Bennett et al. (1999),
Erd\'{e}lyi and
 Carter (2006), and Karami and Barin (2009); b) in $R_1 < r <
R$, within which the resonant layer resides, we use the thin
boundary (TB) approximation in which we assume that the
inhomogeneous layer and the dissipative layer coincide (see Goossens
et al. 2009). This enables us to avoid solving the non-ideal MHD
equations in the inhomogeneous layer. The jumps across the resonant
layer are found by the prescriptions of Sakurai et al. (1991a); c)
substituting the solutions (a) in jump conditions (b) gives an
analytical expression for a dispersion relation to be solved for the
frequencies and the damping rates.

\subsection{Interior and Exterior Solutions}
Following Bennett et al. (1999), Erd\'{e}lyi and
 Carter (2006), and Karami and Barin (2009),
in the absence of dissipations, taking time derivative of Eq.
(\ref{mhd2}) and substituting for $\partial \delta
{\mathbf{v}}/\partial t$ from Eq. (\ref{mhd1}), the resulting
equation yields to Bessel's equation for the Eulerian perturbation
in total pressure $\delta P_{\rm T}$ as
\begin{eqnarray}
\Big[\frac{{\rm d}^2}{{\rm d}r^2}+\frac{1}{r}\frac{{\rm d}}{{\rm
d}r}-\Big(\frac{m^2}{r^2}+m_{\rm i}^2\Big)\Big]{\delta P_{\rm
T}}=0,\label{mhd4}
\end{eqnarray}
where
\begin{eqnarray}
\delta P_{\rm T}=\delta
p+\frac{\mathbf{B}\cdot\delta\mathbf{B}}{4\pi},\label{mhd5}
\end{eqnarray}
and
\begin{eqnarray}
m_{\rm i}^2  = k_z^2 \left[ {1 - \frac{{A^2 \omega _{A_{\rm i}}^2
}}{{\pi \rho_{\rm i} (\omega ^2  - \omega _{A_{\rm i}}^2 )^2 }}}
\right],\label{mhd6}
\end{eqnarray}
\begin{eqnarray}
\omega _{A_{\rm i}}  = \frac{1}{{\sqrt {4\pi \rho_{\rm i} } }}(mA
+ k_z B_z )\label{mhd7}.
\end{eqnarray}
Note that Eq. (\ref{mhd4}) is valid only outside the inhomogeneous
layer. Equation (\ref{mhd4}) is same as the result exactly derived
by Bennett et al. (1999), and Erd\'{e}lyi and Carter (2006). Also
Eq. (\ref{mhd4}) in its general form for a compressible plasma was
derived by Erd\'{e}lyi and Fedun (2007). Note that subscripts $i$
(which correspond to the internal region) are replaced by $e$
corresponding to the external region. Solutions of Eq. (\ref{mhd4})
for the interior region ($r<R_1$) are:
\begin{eqnarray}
\delta P_{\rm T}  = \left\{\begin{array}{lll}
I_{\rm m} (m_{\rm i} r),&m_{\rm i}^2  > 0,&\rm surface~waves, \\
J_{\rm m} (n_{\rm i} r),&n_{\rm i}^2  =  - m_{\rm i}^2  > 0,&\rm body~waves, \\
 \end{array}\right.\label{mhd8}
\end{eqnarray}
where  $J_{\rm m}$ and $I_{\rm m}$ are Bessel and modified Bessel
functions of the first kind, respectively. In the exterior region
($r>R$), the waves should be evanescent. Solutions are
\begin{eqnarray}
\delta P_{\rm T}  =K_{\rm m} (m_{\rm e} r),&m_{\rm e}^2
> 0, \label{mhd9}
\end{eqnarray}
where $K_{\rm m}$ is the modified Bessel function of the second
kind. One must note that from Eq. (\ref{mhd6}), solutions
(\ref{mhd9}) are satisfied only for the frequencies $\omega^2>\omega
_{A_{\rm e}}^2+\frac{A\omega _{A_{\rm e}}}{\sqrt{\pi \rho_{\rm
e}}}$. Our numerical results in Sect. 3 confirms that this condition
is hold for the obtained frequencies.
\subsection{Connection Formulae, Dispersion Relation and Damping}
According to the connection formulae given by Sakurai et al.
(1991a), the jump across the boundary (resonance layer) for
$\xi_r=-\delta v_{r}/{\rm i}\omega$ and $\delta P_{\rm T}$ is
\begin{eqnarray}
\left[ {\xi_r } \right] =  - {\rm i}\pi \frac{1}{|\Delta|
}\frac{{g_B }}{{\rho B^2 }}C_{\rm A},\label{mhd10}
\end{eqnarray}
\begin{eqnarray}
\left[ {\delta P_{\rm T} } \right] =  - {\rm i}\pi
\frac{1}{|\Delta| }\frac{{AB_z F}}{{2\pi \rho B^2}}C_{\rm
A},\label{mhd11}
\end{eqnarray}
where
\begin{eqnarray}
C_{\rm A}&=& g_{\rm B} \delta P_T  - \frac{{AB_z F}}{{2\pi
}}\xi_r,\nonumber\\ \Delta&=& \frac{{\rm d}}{{{\rm
d}r}}\Big(\omega ^2 - \omega _{\rm A}^2 (r)\Big)\Big|_{r=r_{\rm
A}} = \frac{{\rho_{\rm e} - \rho_{\rm i} }}{{a \rho (r_{\rm A}
)}}\omega _{\rm A}^2 (r_{\rm A}),\nonumber\\ F&=& mA + k_z B_z,\nonumber\\
g_{\rm B}&=&\frac{mB_z}{r}-k_zAr.\label{mhd12}
\end{eqnarray}
Note that $a=R-R_1$ is the thickness of the inhomogeneous layer and
$R_1<r_{\rm A}<R$ is the radius at which the singularity occurs.
Davila (1987) showed that in the resonance absorption, however, the
damping rate is independent of the dissipation coefficient values.
But the resonance layer width scales as $\delta_{\rm
A}=[|\frac{\omega}{\Delta}|(\frac{\nu}{\rho}+\frac{c^2}{4\pi\sigma})]^{1/3}$.
Estimates for typical coronal values suggest that resonant layers
have thicknesses from 0.3 km to 250 km (Davila 1987). According to
Ofman et al. (1994), and Erd\'{e}lyi and Goossens (1994, 1995) for
resonant absorption the main contribution comes not from bulk
viscosity but from shear viscosity. If we use the Reynolds
${\mathcal{R}}=\Big(\frac{{\rm R}^2\rho_{\rm
i}}{\nu}\Big)/\Big(\frac{2\pi {\rm R}}{v_{A_{\rm i}}}\Big)=560$ and
Lundquist $S=\Big(\frac{4\pi\sigma {\rm
R}^2}{c^2}\Big)/\Big(\frac{2\pi {\rm R} }{v_{A_{\rm i}}}\Big)=10^4$
numbers given by Ofman et al. (1994), and taking $L=109\times 10^3$
km, $R/L=0.01$, $a/R=0.08$, $\rho_{\rm e}/\rho_{\rm i}=0.1$, and
interior Alfv\'{e}n velocity $v_{A_{\rm i}}=2000$ km s$^{-1}$ for a
typical coronal loop, then one can get $\delta_{\rm A}\simeq 85$ km
which is very close to the thickness of the inhomogeneous layer
$a\simeq 87$ km. Therefore we can use the thin boundary
approximation which assumes that the thickness of the resonance
layer is the same as the inhomogeneous layer width (see Goossens et
al. 2009).

Substituting the solutions of Eqs. (\ref{mhd8})-(\ref{mhd9}) in
jump conditions gives the dispersion relation as
\begin{eqnarray}
d_0 (\tilde \omega ) + d_1 (\tilde \omega ) = 0,\label{mhd13}
\end{eqnarray}
where $\tilde{\omega}=\omega-{\rm i}\alpha$, $\alpha$ is damping
rate and
\begin{eqnarray}
d_0 (\tilde \omega ) = \frac{{m_{\rm i} RI'_{\rm m}(m_{\rm i}
R_1)}}{{I_{\rm m}(m_{\rm i} R_1)}} + \frac{{mFAR}}{{2\pi R_1
D_{\rm i} }} - \frac{{\lambda_{\rm e} }}{{\lambda_{\rm i}
}}\Big(\frac{{m_{\rm e} RK'_{\rm m}(m_{\rm e} R)}}{{K_{\rm
m}(m_{\rm e} R)}} + \frac{{mFA}}{{2\pi D_{\rm e}
}}\Big),\label{mhd14}
\end{eqnarray}
\begin{eqnarray}
d_1 (\tilde \omega ) = \frac{{{\rm i}\pi }}{{|\Delta|\rho (r_{\rm
A} )B^2 (r_{\rm A} )}}\frac{{C_{\rm A} }}{{I_{\rm m}(m_{\rm i} R_1
)}}\Big\{\frac{{g_{\rm B} R}}{{\lambda_{\rm i} }} +
\frac{{\lambda_{\rm e}}}{{\lambda_{\rm i}}}\frac{{FAB_z }}{{2\pi
}}\Big(\frac{{m_{\rm e} RK'_{\rm m}(m_{\rm e} R)}}{{K_{\rm
m}(m_{\rm e} R)}} + \frac{{mFA}}{{2\pi D_{\rm e}
}}\Big)\Big\},\label{mhd15}
\end{eqnarray}
with
\begin{eqnarray}
\lambda _{\rm j}&=& \frac{{D_{\rm j} }}{{D_{\rm j}^2  - 4(FA/4\pi
)^2}},
\nonumber\\
D_{\rm j}&=&\frac{{F^2 }}{{4\pi }} - \rho_{\rm j} \omega
^2,\label{mhd16}
\end{eqnarray}
where (j) stands for (i) or (e). Also a prime on $I_{\rm m}$ and
$K_{\rm m}$ indicates a derivative with respect to their
appropriate arguments. The results for the body waves are the same
as Eqs. (\ref{mhd14})-(\ref{mhd15}), except that $I_{\rm m}$ is
replaced by $J_{\rm m}$ everywhere.

Equations (\ref{mhd10})-(\ref{mhd11}) show that when the twist is
absent, i.e. $A=0$, the total Eulerian pressure will be continuous
and only $\xi_r$ jumps across the boundary which is in agreement
with Sakurai et al. (1991a).

For the surface modes, solving Eq. (\ref{mhd13}) yields one mode
whose frequency exists between $\omega_{\rm A_i}$ and $\omega_{\rm
A_e}$. But for the body modes, there is an infinite band of body
modes with the frequencies centered around $\omega_{\rm A_i}$. The
existence of an infinite set of body waves in the presence of twist
has been already introduced by Bennett et al. (1999).

Note that the distinction between surface and body waves is not so
important when we study the kink waves in slender magnetic tubes.
Recently Goossens et al. (2009) concluded in the thin tube
approximation, the kink MHD waves do not care about propagating
(body wave) or evanescent (surface wave) behavior in the internal
part of the flux tube. Although under coronal condition they are
formally the body waves, they have typical properties of the surface
waves. For example, the radial displacement takes its maximum value
at the tube boundary. This is why Ruderman and Roberts (2002)
suggested to call these waves the global kink waves. Also
Erd\'{e}lyi and Carter (2006) showed that under coronal conditions,
for longer wavelength values the sausage surface mode only exists
for smaller amounts of twist in the incompressible plasma.

The numerical solution of the dispersion relation (\ref{mhd13})
yields to frequencies $\omega_{nml}$ and damping rates
$\alpha_{nml}$, which are characterized by a trio of wavenumbers
($n$, $m$ and $l$) that actually count the number of nodes or
antinodes along $r$-, $\phi$- and $z$-directions, respectively.

\section{Numerical Results}
As typical parameters for a coronal loop, we assume $L=109\times
10^3$ km, $R/L=0.01$, $a/R=0.08$, $\rho_{\rm e}/\rho_{\rm i}=0.1$,
$\rho_{\rm i}=2\times 10^{-14}$ gr cm$^{-3}$, $B_{z}=100$ G. For
such a loop one finds $v_{A_{\rm i}}=2000$ km s$^{-1}$, $\omega_{\rm
A_i}:=\frac{v_{\rm A_i}}{ L}\simeq 0.02$ rad s$^{-1}$.

\begin{figure}
\center \includegraphics{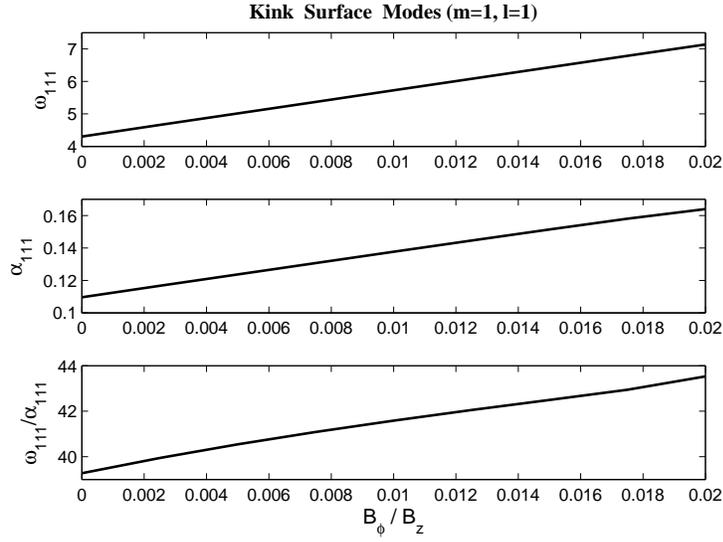}
      \vspace{7.2cm}
      \caption[]{Frequency of the fundamental kink ($m=1$) surface mode, its damping
rate and the ratio of the oscillation frequency to the damping
rate versus the twist parameter, $B_{\phi}/B_z$. The loop
parameters are: $L=109\times 10^3$ km, $R/L=0.01$,  $a/R=0.08$,
$\rho_{\rm e}/\rho_{\rm i}=0.1$,
 $\rho_{\rm i}=2\times 10^{-14}$ gr
cm$^{-3}$, $B_{z}=100$ G. Both frequencies and damping rates are
in units of the interior Alfv\'{e}n frequency, $\omega_{\rm
A_i}\simeq 0.02{\rm~rad~s^{-1}}$.}
         \label{m1l1-surface-wa}
   \end{figure}

\begin{figure}
\center \includegraphics{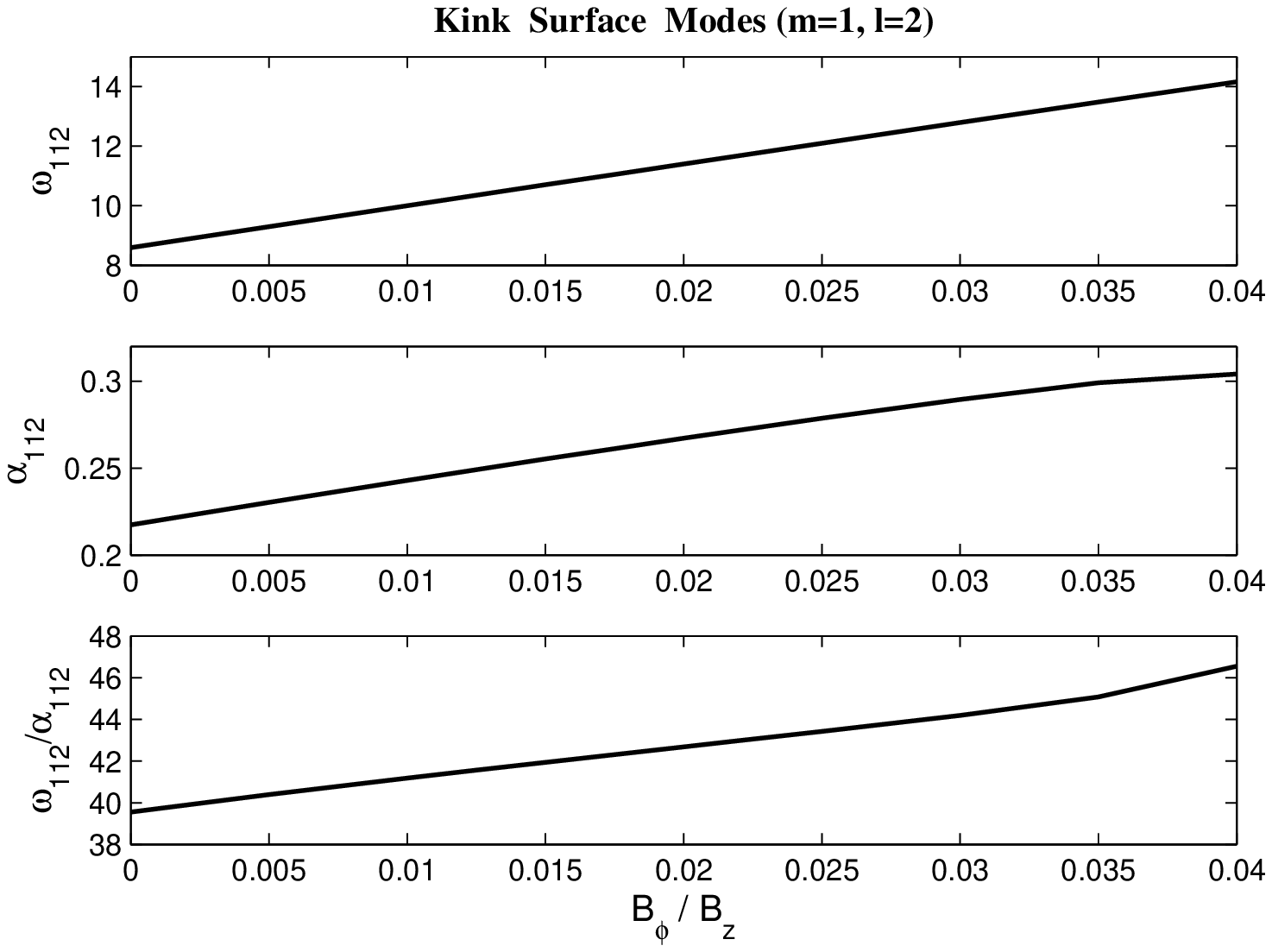}
      \vspace{4.5cm}
      \caption[]{Same as Fig. \ref{m1l1-surface-wa}, for the first-overtone kink ($m=1$) surface modes.}
         \label{m1l2-surface-wa}
   \end{figure}

\begin{figure}
\center \includegraphics{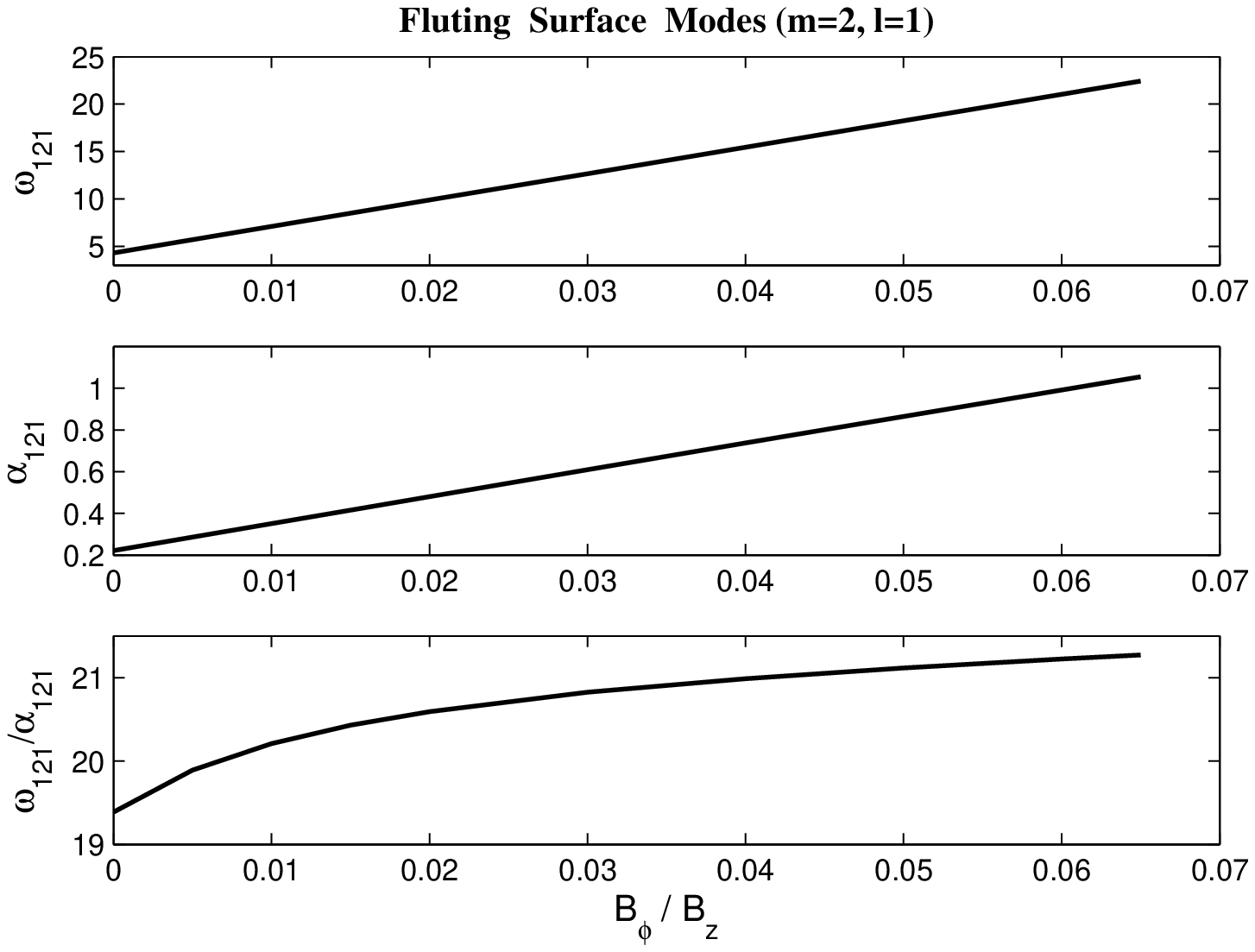}
      \vspace{8.6cm}
      \caption[]{Same as Fig. \ref{m1l1-surface-wa}, for the fundamental fluting ($m=2$) surface modes.}
         \label{m121-surface-wa}
   \end{figure}

\begin{figure}
\center \includegraphics{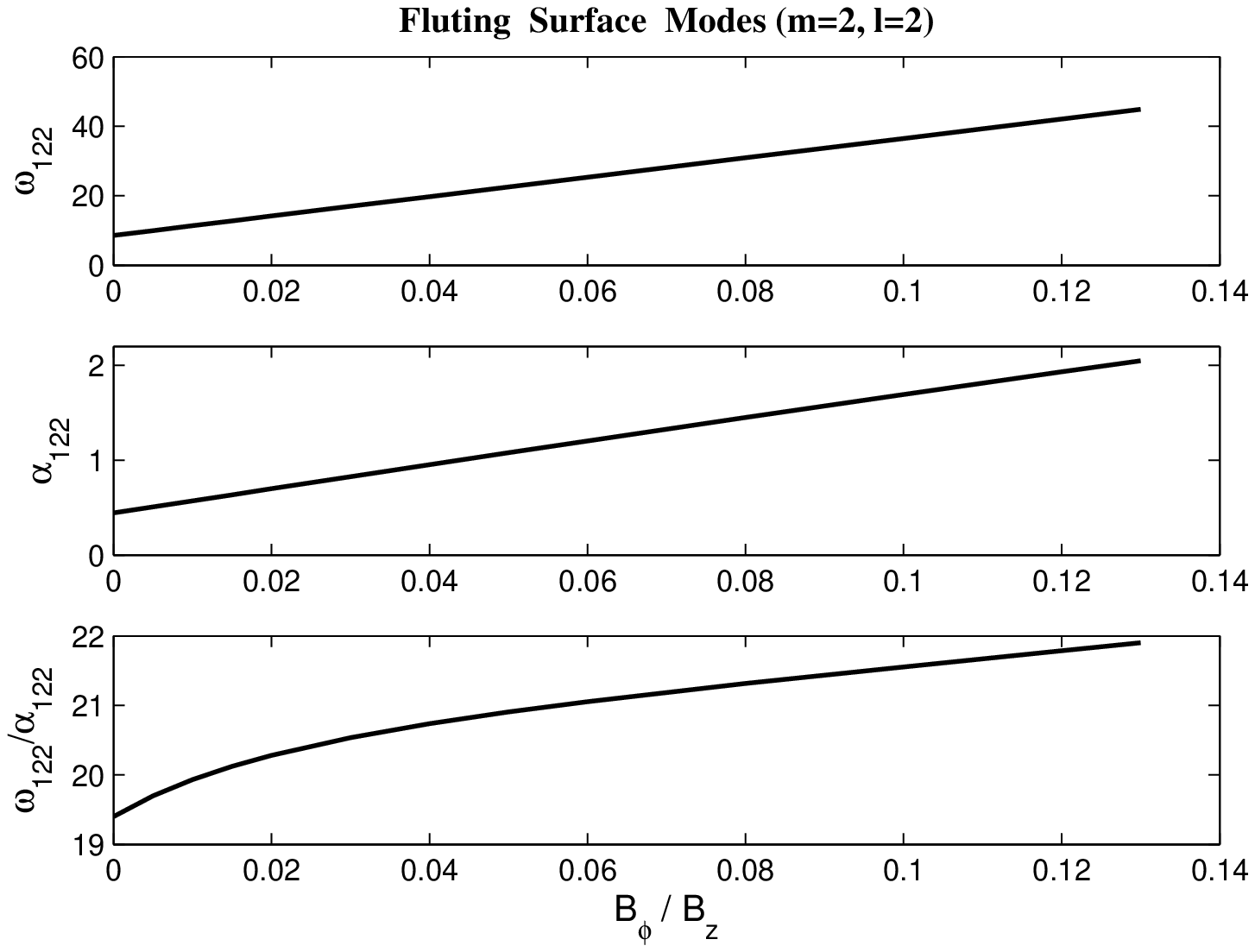}
      \vspace{8.6cm}
      \caption[]{Same as Fig. \ref{m1l1-surface-wa}, for the first-overtone fluting ($m=2$) surface modes.}
         \label{m122-surface-wa}
   \end{figure}

\begin{figure}
\center \includegraphics{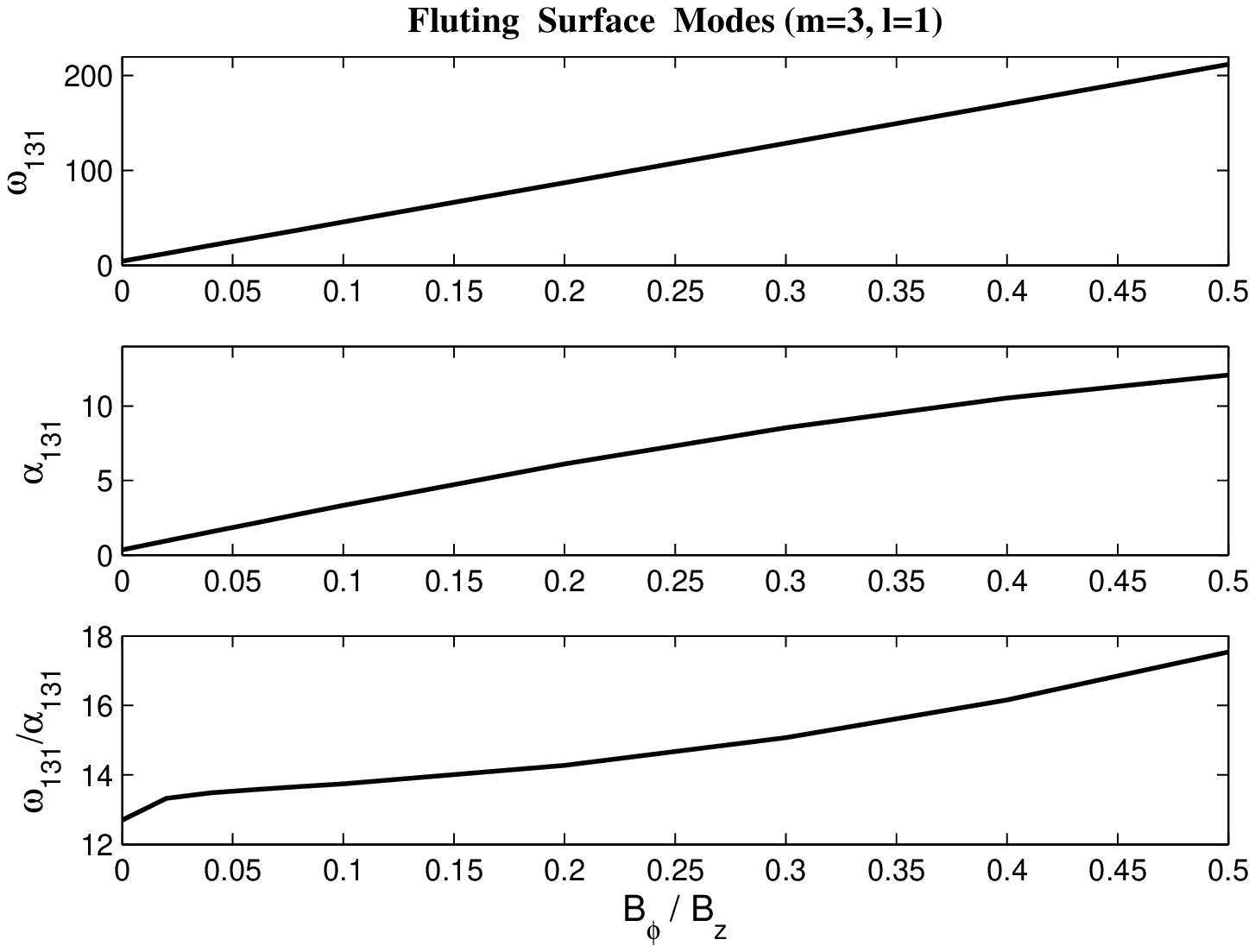}
      \vspace{8.6cm}
      \caption[]{Same as Fig. \ref{m1l1-surface-wa}, for the fundamental fluting ($m=3$) surface modes.}
         \label{m131-surface-wa}
   \end{figure}

\begin{figure}
\center \includegraphics{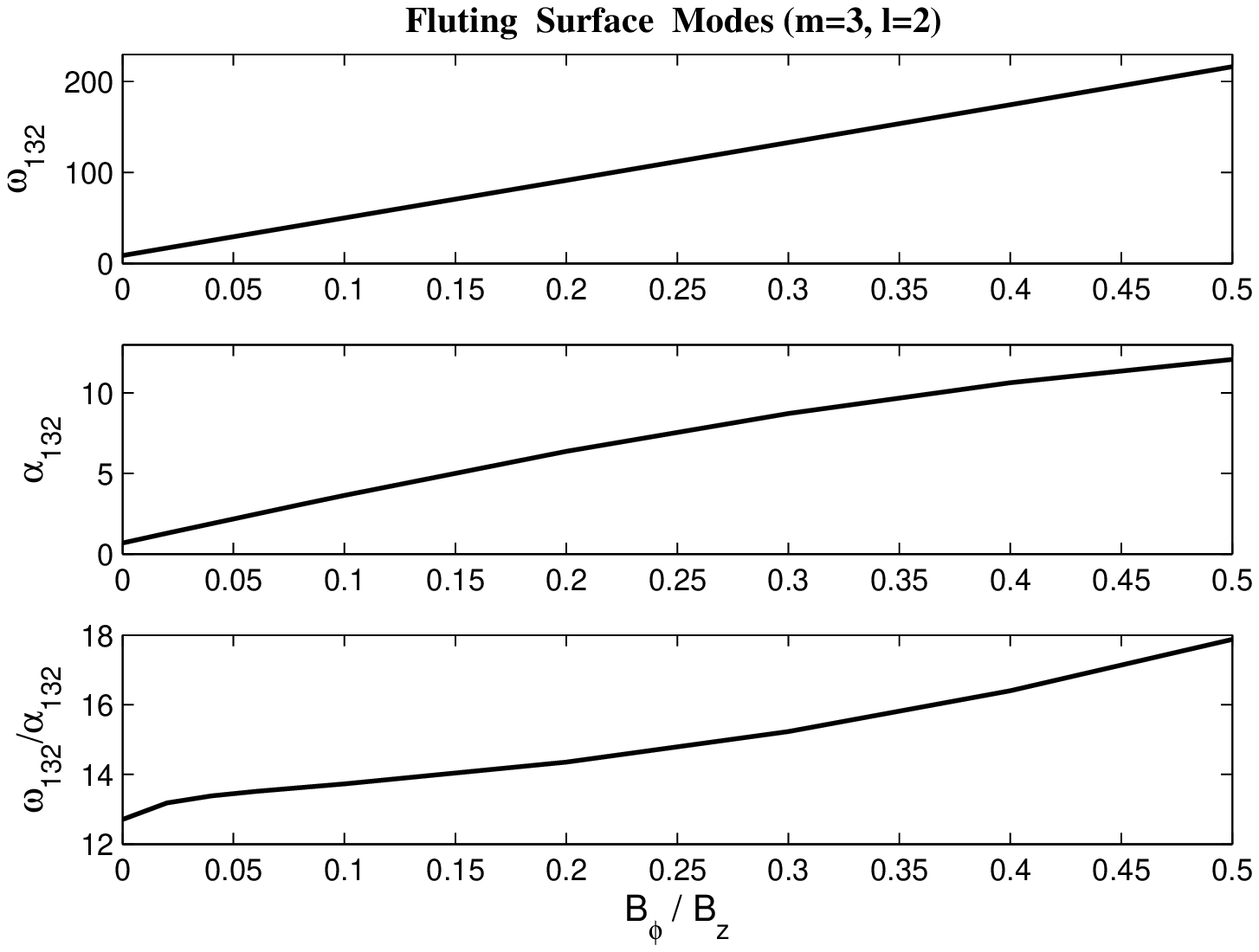}
      \vspace{8.6cm}
      \caption[]{Same as Fig. \ref{m1l1-surface-wa}, for the first-overtone fluting ($m=3$) surface modes.}
         \label{m132-surface-wa}
   \end{figure}

\begin{figure}
\center \includegraphics{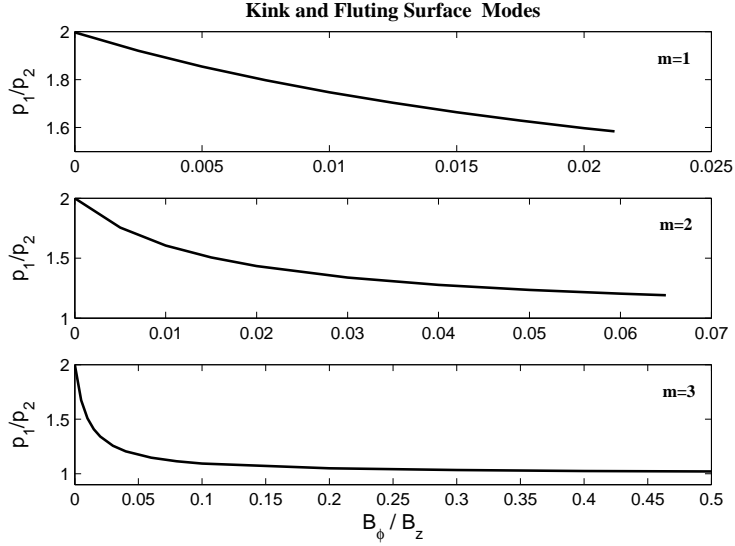}
      \vspace{8.6cm}
\caption[]{The period ratio $P_1/P_2$ of the fundamental and its
first-overtone surface modes versus the twist parameter
  for kink ($m=1$) and fluting ($m=2,3$) modes. Auxiliary parameters as in Fig.
\ref{m1l1-surface-wa}.}
         \label{surface-p1p2}
   \end{figure}

\begin{figure}
\center \includegraphics{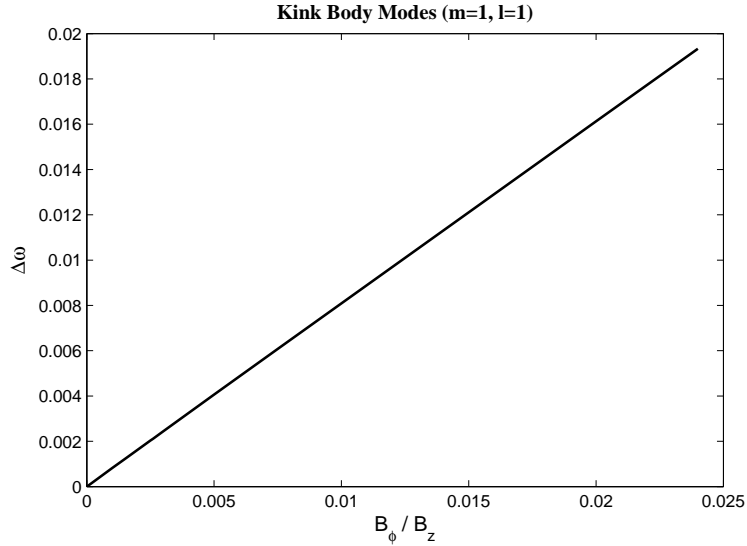}
      \vspace{8.6cm}
\caption[] {Frequency bandwidth of the fundamental kink (m=1) body
modes versus the twist parameter for the relative inhomogeneous
layer width $a/R=0.08$. Auxiliary parameters as in Fig.
\ref{m1l1-surface-wa}.}
         \label{bandwith}
   \end{figure}

The effects of twisted magnetic field on both the frequencies
$\omega$ and damping rates $\alpha$ are calculated by numerical
solution of the dispersion relation, i.e. Eq. (\ref{mhd13}). The
results are displayed in Figs. \ref{m1l1-surface-wa} to
\ref{bandwith}. Figures \ref{m1l1-surface-wa} to
\ref{m132-surface-wa} show the frequencies, damping rates and also
the ratio of the oscillation frequency to the damping rate of the
fundamental and first-overtone $l=1,2$ kink $(m = 1)$ and fluting
$(m=2,3)$ surface modes versus the twist parameter,
$B_{\phi}/B_z:=\frac{AR}{B_z}$ for relative inhomogeneous layer
width $a/R=0.08$. According to Carter and Erd\'{e}lyi (2008), the
kink $(m = 1)$ speed is independent of the sound speed. Hence, the
kink modes are highly incompressible, and this makes the limit of
incompressibility be of great interest, not just for wave studies in
the deeper part of the solar atmosphere (e.g. where the plasma-beta
is high) but can also be directly applicable from the lower solar
atmosphere to the corona. For all other modes, $(m \neq 1)$,
compressibility may be necessary.

Figures \ref{m1l1-surface-wa} to \ref{m132-surface-wa} reveal that:
i) the frequencies, the damping rates and the ratio $\omega/\alpha$
increase when the twist parameter increases. The result of $\omega$
is in good agreement with that obtained by Carter and Erd\'{e}lyi
(2008). Also the behavior of $\omega$ and $\alpha$ versus the twist
parameter are in good concord with that obtained by Karami and Barin
(2009). But, the numerical values of our damping rates for $m=$1, 2
and 3 are six, seven and eight orders of magnitude, respectively,
greater than those obtained by Karami and Barin (2009). This is due
to existence of the resonant absorption which is absent in their
work. Note that the dissipative processes in Karami and Barin (2009)
depend on the Reynolds ${\mathcal{R}}$ and Lundquist $S$ numbers.
They obtained the damping rates with ${\mathcal{R}}=560$ and
$S=10^4$ given by Ofman et al. (1994). But in the resonance
absorption, the damping rate is independent of the dissipation
coefficient values and only the resonance layer width depends on
${\mathcal{R}}$ and $S$ (see Davila 1987). ii) For $m=1$ with
increasing $B_{\phi}/B_z$, the frequencies, damping rates and their
ratio, for instance $\omega_{111}$, $\alpha_{111}$ and
$\omega_{111}/\alpha_{111}$, increase $\simeq$8-66, $\simeq$6-50 and
$\simeq$2-11 percent, respectively, compared with an untwisted loop.
iii) For a given $m$ and $B_{\phi}/B_z$, when the longitudinal mode
number, $l$, increases, both the frequencies and damping rates
increase. But the ratio $\omega/\alpha$: for $m=1$, for the twists
lower than 0.003 increases and then decreases; for $m=2$, always
decreases; but for $m=3$, for the twists lower than 0.115 decreases
and then increases. iv) For a given $l$ and $B_{\phi}/B_z$, when the
azimuthal mode number, $m$, increases, the frequencies and damping
rates increase but the ratio $\omega/\alpha$ decreases.

Here in our calculations, the sausage modes ($m=0$) are absent.
Because following Edwin and Roberts (1983), and Roberts, Edwin, and
Benz (1984), the sausage modes have a cut-off longitudinal
 wave number $l_{\rm c}$,
and they are only expected in fat and dense loops. According to
Aschwanden (2005) the longitudinal wave number cut-off $l_{\rm c}$
for sausage-mode oscillations is expressed as a requirement of the
loop length-to-width ratio $L/(2R)$ as a function of the density
contrast $\rho_{\rm e}/\rho_{\rm i}$ between the external and
internal loop densities. For instance, for a typical active region,
loops which have a density contrast of the order of $\rho_{\rm
e}/\rho_{\rm i}\approx0.1-0.5$ would be required to have
length-to-width ratios of $L/(2R)\approx 1-2$.

The period ratio $P_1/P_2$ of the fundamental and first-overtone,
$l=1,2$ modes of both the kink ($m=1$), and fluting ($m=2,3$)
surface waves versus the twist parameter is plotted in Fig.
\ref{surface-p1p2}. Figure shows that for both kink and fluting
modes, the period ratio $P_1/P_2$ decreases when the twist parameter
increases. For instance, $P_1/P_2$ decreases from 2 (for untwisted
loop) and approaches below 1.6, 1.2 and 1.1 for $m$ = (1, 2 and 3),
respectively, with increasing the twist parameter. Note that when
the twist is zero, the diagrams of $P_1/P_2$ do not start exactly
from 2. This may be caused by the radial structuring $\rho_{\rm
e}\neq\rho_{\rm i}$. But for the selected thin tube with $R/L=0.01$,
this departure is very small, $O(10^{-3})$ for kink ($m=1$) and
$O(10^{-4})$ for fluting ($m=2, m=3$) modes, and does not show
itself in the diagrams (see McEwan et al. 2006). Figure
\ref{surface-p1p2} clears that for kink modes $(m=1)$, the ratio
$P_1/P_2$ is 1.821 when $B_{\phi}/B_z$=0.0065, and is 1.584 for
$B_{\phi}/B_z$=0.0212. These are in good agreement with the period
ratios observed by Van Doorsselaere et al. (2007), 1.82$\pm$0.08 and
1.58$\pm$0.06, respectively, deduced from the observations of TRACE.
Van Doorsselaere et al. (2007) considered only the effect of density
stratification and neglected the effect of loop expansion. It has
been shown by Verth and Erd\'{e}lyi (2008) that failure to take
account of the flux tube expansion causes the density scale height
to be overestimated by a factor of approximately 2. Verth et al.
(2008) considered both effects of the longitudinal density
stratification and magnetic field expansion and estimated the
correct coronal
 density scale height. They reported a period ratio of
1.54 deduced from the observations of TRACE for the kink modes. Here
in our numerical calculations, if we take $B_{\phi}/B_z$=0.0255 and
$\rho_{\rm e}/\rho_{\rm i}=0.05$ then we can obtain the period ratio
$P_1/P_2$=1.539 for the kink modes which is in good agreement with
that reported by Verth et al. (2008). This shows that the magnetic
twist can achieve deviations from 2 of the same order of magnitude
as in the model of Verth et al. (2008) containing the longitudinal
density stratification and magnetic field expansion. Note that for
the kink modes, the only period ratio obtained by Karami and Barin
(2009) was 1.882 with $B_{\phi}/B_z$=0.0065. Therefore considering
the magnetic twist in presence of the resonant absorption not only
improved the period ratio estimated by Karami and Barin (2009) but
also justified another period ratio, 1.54, reported by Verth et al.
(2008).

Figure \ref{bandwith} displays the frequency band width,
$\Delta\omega$, including infinite set of the fundamental kink
$(m=1)$ body modes versus the twist parameter. Figure \ref{bandwith}
presents that $\Delta\omega$ increases when the twist parameter
increases. This is in good agreement with the result obtained by
Carter and Erd\'{e}lyi (2008), and Karami and Barin (2009).

\section{Conclusions}
Resonant absorption of standing MHD surface and body waves in
coronal loops in the presence of the twisted magnetic field is
studied. To do this, a typical coronal loop is considered as a
straight cylindrical incompressible flux tube with magnetic twist
that undergoes a radial density structuring. Under the thin tube
approximation, the thickness of the inhomogeneous layer is same as
the resonance layer width. Hence, the radial structuring is assumed
to be a linearly varying density profile in the inhomogeneous layer.
Using the relevant connection formulae, the dispersion relation is
obtained and solved numerically for obtaining both the frequencies
and damping rates of the fundamental and first-overtone kink and
fluting modes. Our numerical results show that

i) the frequencies and damping rates as well as the ratio of the
oscillation frequency to the damping rate of both the kink $(m = 1)$
and fluting $(m=2,3)$ surface waves increase when the twist
parameter increases. For $m=1$ with increasing $B_{\phi}/B_z$, the
ratio $\omega_{111}/\alpha_{111}$ changes from 39.3 to 43.5 which
approximately one order of magnitude is greater than the ratio
reported by Nakariakov et al. (1999), Wang and Solanki (2004), and
Verwichte et al. (2004) deduced from the TRACE data;

ii) the period ratio $P_1/P_2$, for both the kink ($m=1$) and
fluting ($m=2,3$) surface modes is lower than 2 (for an untwisted
loop) in the presence of the twisted magnetic field. The result of
$P_1/P_2=1.821,1.539$ for kink modes is in agreement with the
TRACE observations;

iii) Frequency bandwidth of the fundamental kink ($m=1$) body
modes increase when the twist parameter increases.

Note that the main differences between the present work and the work
of Karami and Barin (2009) can be summarized as follows: i) in our
model, the magnetic field was assumed to be uniformly twisted inside
and outside the tube. Whereas they considered only a uniformly
twisted magnetic annulus inside the loop. ii) They considered the
weak damping due to viscous and resistive dissipations which cannot
justify the rapid damping of the coronal loops observed by
Nakariakov et al. (1999), and Wang and Solanki (2004). In the case
one includes the resonant absorption, the rapid damping can be
justified. iii) Our model could justify the period ratios of kink
modes reported by Van Doorsselaere et al. (2007), i.e.
$P_1/P_2=$1.82, 1.58, and Verth et al. (2008), i.e. $P_1/P_2=$1.54,
whereas Karami and Barin (2009) only justified $P_1/P_2=$1.82.

\section*{Acknowledgments}
The authors thank the unknown referee for very valuable comments.
The authors also thank Dr. H. Safari for a number of useful
discussions. This work was supported by the Department of Physics,
University of Kurdistan, Sanandaj, Iran.


\clearpage

\end{document}